# Nano-Confinement Effects on Liquid Pressure

An Zou & Shalabh C. Maroo*
Department of Mechanical and Aerospace Engineering, Syracuse University, Syracuse, NY 13244
* Corresponding author: scmaroo@syr.edu

## ABSTRACT:

In this work, molecular dynamics simulations are performed to estimate the equilibrium pressure of liquid confined in nanopores. The simulations show that pressure is highly sensitive to the pore size and can significantly change from absolute positive to absolute negative values for a very small (0.1 nm) change in the pore size. The contribution from the solid-liquid interaction always dominates the pressure in the first liquid layer adjacent to the surface and the sensitiveness of pressure on the pore size is dependent on the atom distribution in the liquid layers. A surface influence number $S$ is introduced to quantitatively characterize the degree of the confinement. At constant system temperature, the $S$ number decreases with increasing pore size based on a power law function. In nanopores with large $S$ number, the pore liquid pressure is found to be independent of bulk liquid pressure, whereas in nanopores with small $S$ number, the pore pressure is dependent and increases with bulk pressure.

## I. INTRODUCTION

Fluids confined in channels or pores at nanoscale are of great importance, and can be found in a wide variety of natural and engineering systems, such as water confined in cells of living organisms,[1] transpiration,[2, 3] high heat flux removal for electronics cooling,[4-7] nanofluidic devices for desalination,[8-10] drug delivery,[11, 12] etc. Nanoscale confined fluids have shown physical, chemical, and thermodynamics properties dramatically different from their bulk properties due to the presence of strong solid-liquid intermolecular interactions.[13-15] A comprehensive knowledge linking the molecular-level characteristics and the macroscopic fluid properties is of great significance to design novel nanoscale structures/devices for desired applications, as well as to better understand our natural systems. Fluid pressure in a confined environment has been a topic of interest due to the aforementioned reasons.[16-19] In this paper, we focus on the equilibrium pressure of liquid confined in nanopores of decreasing sizes to the extent that only solid-liquid interface exists without any bulk fluid.

At the solid-liquid interface, it is well-known that liquid layer structuring occurs,[20-22] and the liquid atoms adopt a configuration based on the solid atoms lattice structure and spacing.[23-26] The structured liquid layers on the surface are usually associated with high density and high pressure;[27-32] however, absolute negative pressures can occur in low density layers depending on the dimension of the channels/pores.[18, 33, 34] Here, we report a fundamental molecular dynamics (MD) study of the atypical pressure of confined fluids in hydrophilic nanopores connected to bulk fluids. Based on atom groups, we differentiate the contributions of solid-liquid and liquid-liquid interactions to the overall pressure by introducing a recurring ghost-step in the simulations where liquid-liquid interaction is artificially set as zero; the atom trajectories at each time step are obtained as in a typical MD simulation where all intermolecular interactions are included. We discuss confinement effects on the structure of the first liquid layer adjacent to the solid surface and its resulting positive/negative pressure along with the effects of bulk pressure on pore pressure. A surface influence number is introduced to quantitatively characterize the degree of confinement of a nanopore.

## II. METHODS

### A. MD Simulations

Figure 1 shows a side view of the 3D simulation domain, which includes a nanopore in thermodynamic equilibrium with bulk liquid/vapor. The nanopore is formed between two 5 nm long parallel hydrophilic surfaces with distance of $W$, which defines the nanopore characteristic length. The hydrophilic (HL) surface consists of five layers of FCC <111> plane atoms. The bulk liquid is a 5.2 nm thick continuous film, which is contained by a 2.2 nm thick hydrophobic (HP) surface at the bottom. The same HP atoms fill the space between HL surfaces and domain boundaries to serve as upper-side boundaries for bulk liquid and the lower-side boundaries for the vapor. All MD simulations were run in LAMMPS.[35]

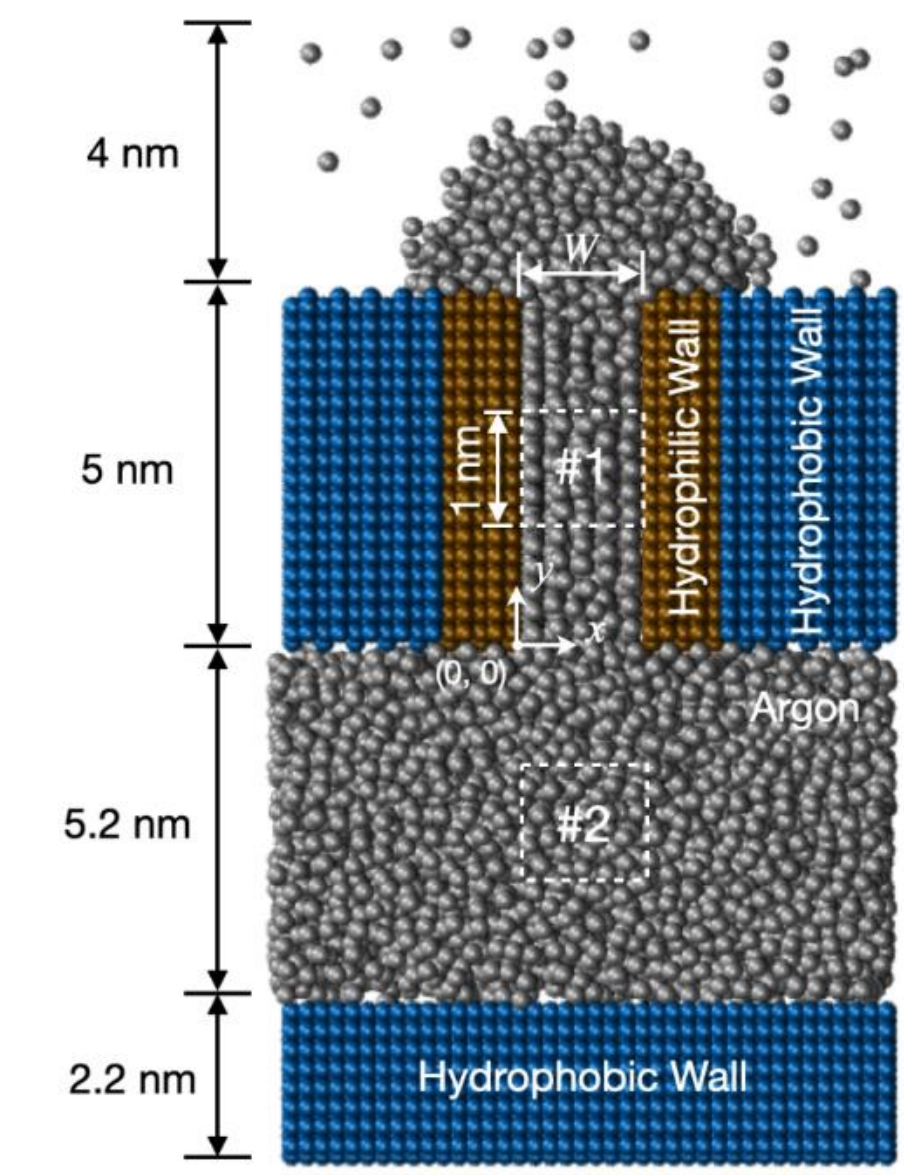

*Figure 1: Molecular Dynamics simulation domain of a nanopore connected to bulk liquid.*

Argon (Ar) fluid was chosen in current work as its thermodynamic properties, statistically obtained from MD, are in good agreement with experimental data over the entire temperature range using 12-6 Lennard-Jones (L-J) potential.[36-38] All atomic interactions are governed by 12-6 L-J potential with a force smoothing applied between the inner and outer cutoff (Eq. 1), which are fixed as 1.8 nm and 2.0 nm respectively for all atom combinations in current study.

$$\phi = 4\varepsilon\left[\left(\frac{\sigma}{r}\right)^{12} - \left(\frac{\sigma}{r}\right)^{6}\right] \qquad r < r_{in}$$

$$F = C_1 + C_2(r - r_{in}) + C_3(r - r_{in}^2 + C_4(r - r_{in}^3) \quad r_{in} < r < r_c \qquad (1)$$

where $\phi$ is L-J potential; $\varepsilon$ is the depth of the potential well; $\sigma$ is the distance where the potential is zero; $r$ is the distance between two atoms; $r_{in}$ and $r_c$ are inner and outer cutoff respectively; $F$ is the force between two atoms; $C_1$, $C_2$, $C_3$, and $C_4$ are coefficients calculated by LAMMPS for the force varying smoothly from $r_{in}$ to $r_c$.

The HL surface was mimicked by setting $\varepsilon_{Ar-HL}$ as 1.5 times $\varepsilon_{Ar-Ar}$ while $\varepsilon_{Ar-HP}$ was set as 1% of $\varepsilon_{Ar-Ar}$ for the HP surface. Thus, the HP surface only served as a physical barrier without affecting Ar atom dynamics. Table 1 lists the parameters of L-J potentials for all atom combinations. The surface wettability was verified by simulating the spread of a cubic Ar drop on the surface. The drop spread completely on HL surface implying a 0° contact angle while it became a near spherical drop showing a contact angle of ~180° on the HP surface.[39]

*Table 1: 12-6 L-J potential parameters*

| Combination | ε ($10^{-21}$ J) | σ (nm) |
|---|---|---|
| Ar – Ar | 1.67 | 0.3400 |
| Ar - HL | 2.49 | 0.3085 |
| Ar - HP | 0.0167 | 0.3085 |

At start of the simulations, liquid Ar atoms were placed in the bulk and nanopore while vapor Ar atoms were placed above the nanopore (please see supplementary material). The number of Ar atoms in the system were adjusted to obtain a desired bulk pressure (please see supplementary material). A total of 14 cases were run with four different pore sizes (0.9 nm, 1 nm, 2 nm, and 3 nm) and different bulk pressures. For each case, the system was equilibrated at 90 K for 3 ns (600,000 steps) in a canonical NVT ensemble (N is the number of atoms, V is the volume, and T is the temperature) using Nose-Hoover thermostat.[40, 41] During the equilibrium, a convex meniscus formed at the vapor side of the pore, as the contact line advancement was inhibited by the hydrophilic-hydrophobic boundary (Fig. 1).[42] The data (density and pressure) were averaged over each 0.1 ns. The average of last 1 ns (200,000 steps, 10 outputs) was taken as the equilibrium value, and the standard deviation was used for error bars.

**B. Density and Pressure**

In order to obtain the local density and pressure, the domain was divided into bins of 0.05 nm × 0.05 nm in $x$ and $y$ directions. The density distribution was calculated by counting the number of atoms in each bin (Eq. 2). The local pressure was obtained from the summation of normal stresses for each atom in the respective bin (Eq. 3).[43] The data for nanopore was obtained by averaging bin values at the center region with a length of 1 nm (box #1 in Fig. 1) to eliminate any direct influence from the bulk and meniscus regions. A region of W × 1 nm$^2$ (in $x$ and $y$ direction respectively; W is width) was chosen to represent bulk average values (box #2 in Fig. 1) in order to have no effect from any surface. The surfaces forming the nanopore were in the y-z plane; thus the tangential pressure ($P_T$) was obtained by averaging two in-plane components, $P_T = (P_{yy} + P_{zz})/2$; while the normal pressure was obtained from $P_{xx}$ directly ($P_N = P_{xx}$).

$$\rho_i = \frac{N_i M_{Ar}}{N_A V_i} \quad (2)$$

$$P_i = \sum_{k=x,y,z} P_{i,kk} = \frac{1}{3V_i} \sum_{k=x,y,z} \left[ \frac{M_{Ar}}{N_A} v_k^2 - \frac{1}{2} \sum_{n=1}^{N_p} \frac{d_k}{r} \phi'(r) \right] \quad (3)$$

where $N_i$ is number of atoms in $i$th bin, $M_{Ar}$ is molecular mass of Ar, $N_A$ is Avogadro number, $V_i$ is volume of the bin, $v_k$ is the atom velocity at $k$ direction. The first term in summation of Eq. 3 is the kinetic energy contribution $P_{ke}$; while the second term is pairwise energy contribution, $P_{pair}$, where $n$ loops over $N_p$ neighbors of atom, $d_k$ is the distance between two atoms in $k$ direction, and $\phi'(r)$ is the first order derivative of L-J potential with respect to $r$. The constant 1/2 implies that if only one atom of the pair is in $i$th bin, half of the intermolecular force contribution is given to the current bin; total contribution is given to current bin if both atoms are located in the same bin.

### C. Force Separation

In order to differentiate the pressure components based on atom groups, the total force experienced by an Ar atom was separated to solid-liquid (SL) force from surface atoms and liquid-liquid (LL) force from neighboring Ar atoms. The force separation was achieved by running two sets of MD simulations as shown in Fig. 2: one named as "MD-TOT" which included all pairs of interactions to simulate the evolution of the system and generate the total pressure data; and the other denoted as "MD-SL" which only included solid-liquid interaction in order to calculate the corresponding pressure data during an intermediate ghost-step as explained next. In the $i$th loop, the MD-TOT simulation reads the restart file from the $(i-1)$th loop and runs for one step to generate (1) total pressure data $P_{TOT}$ of $i$th step, and (2) a restart file including velocities, positions, etc. of all atoms in system. Then, the MD-SL simulation reads the generated restart file, and runs for 0 step (i.e. ghost-step) to generate solid-liquid interaction contributed pressure $P_{SL}$ data of $i$th step without any atom movement under the new force field; this was achieved by artificially setting the liquid-liquid interaction to 0. After the data processing, the next loop starts to continue the simulation. Starting from a system in equilibrium, this process was repeated for 200,000 times to get the data for 1 ns. In our system, the interactions between pairs of atoms are independent and additive. Thus, the difference between total pairwise pressure $P_{pair,TOT}$ and solid-liquid pairwise pressure $P_{pair,SL}$ is the pairwise pressure contributed by liquid-liquid interaction ($P_{pair,LL} = P_{pair,TOT} - P_{pair,SL}$). Please see details on verification of the force separation process in supplementary material.

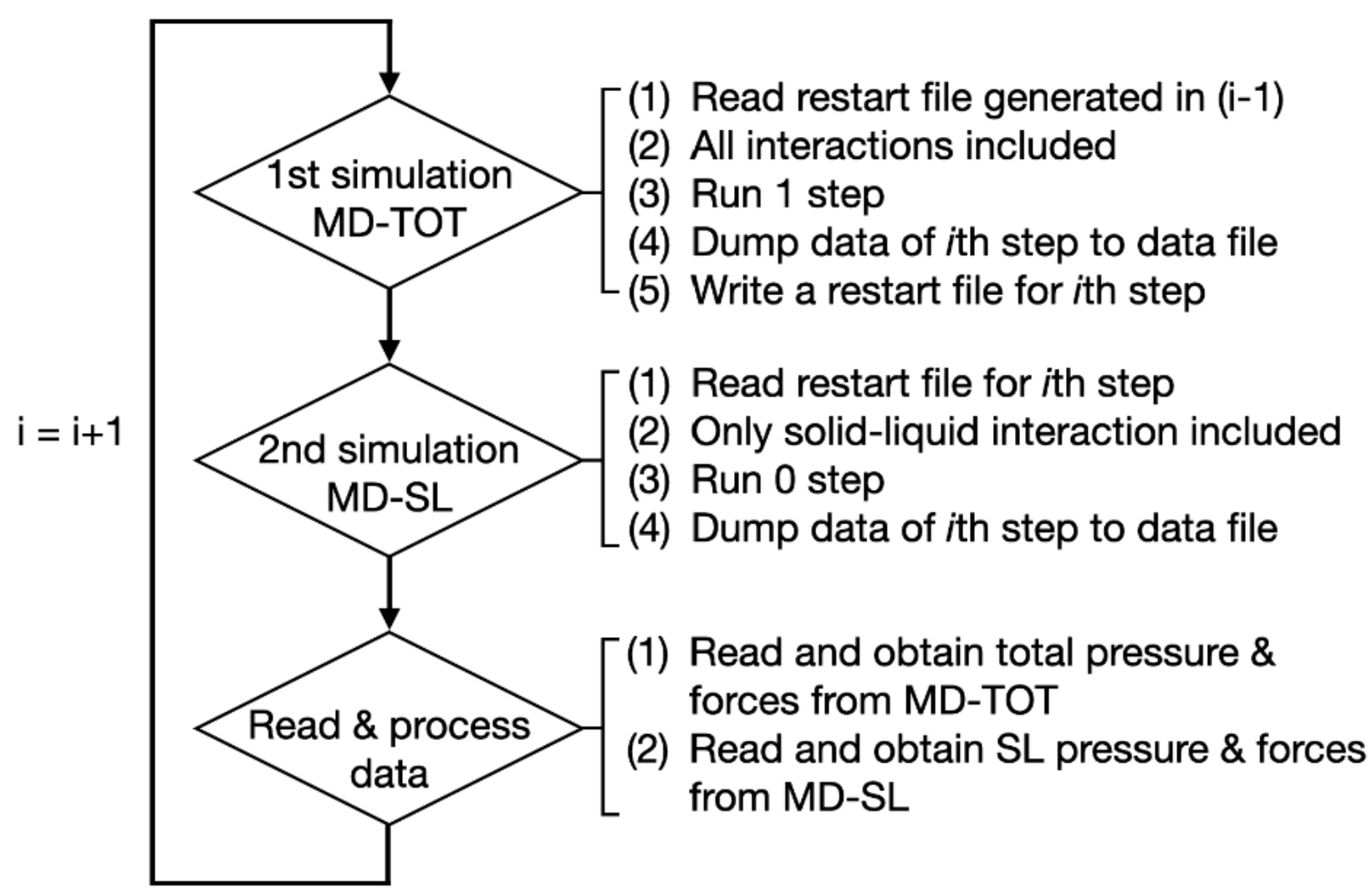

*Figure 2: Flowchart to differentiate pressure contributions based on atom groups.*

## III. RESULTS AND DISCUSSION

### A. Pressure Tensors in Nanopores

The liquid pressure at the solid-liquid interface is expected to be much higher than bulk liquid due to the structuring of liquid atoms in response to solid-liquid interaction.[15–17,22–27] Typically, a thin film on a solid surface has a free liquid-vapor interface (i.e. film is exposed to vapor). However, in a confined nanopore, solid-liquid interface can exist without any bulk liquid or free liquid-vapor interface. In such cases, the liquid pressure is highly dependent on the pore size $W$. Interestingly, for a nanopore with a certain number of liquid layers, a slight increase in $W$ may not create enough space for the formation of an additional layer; nonetheless, as the volume has increased, the pressure decreases or even turns absolute negative. If W is further increased, the pore becomes wide enough to add another liquid layer resulting in a jump to high and positive pressure. Thus this pressure oscillation, between reduced/negative to positive values, repeats periodically with increasing $W$ and has a period similar to L-J diameter of the liquid atoms.[28] In order to estimate the pressure tensors and separate the contributions of solid-liquid and liquid-liquid interactions in determining positive/negative pressures in the nanopore, we first ran two cases with $W$ of 0.9 nm (~$2.65\sigma_{Ar-Ar}$) and 1.0 nm (~$2.94\sigma_{Ar-Ar}$) respectively. Two liquid layers existed in the nanopore for both cases. The bulk pressure was maintained close to the saturation pressure (1.32 atm at 90 K) for both cases (4.52 ± 18.01 atm and 0.95 ± 6.62 atm for 0.9 nm and 1.0 nm pores respectively).

Figures 3 and 4 show the profiles of density and pressure tensors in 0.9 nm and 1.0 nm pores respectively. As expected, the liquid pressure in 0.9 nm pore was high and positive (686.82 ± 23.44 atm); while the pressure was negative in 1.0 nm pore (−451.51 ± 29.49 atm). The density profiles show the occurrence of structured layers of liquid atoms near the surface. The pressure profile follows the density profile as extreme pressure (either positive or negative) occurs near the location of maximum density (Figs. 3b and 4b). Tangential pressure $P_{T,TOT}$ is found to be positive in both cases. However, the normal pressure $P_{N,TOT}$ dominates the total pressure $P_{TOT}$, regardless of $P_{N,TOT}$ being positive or negative. In terms of the type of contribution to the pressure, pairwise energy contribution $P_{pair,TOT}$ dominates in both cases compared to the kinetic energy $P_{ke,TOT}$ component. The pressure contributed by $P_{ke,TOT}$ is always positive due to the above-zero temperature (Figs. 3c and 4c).

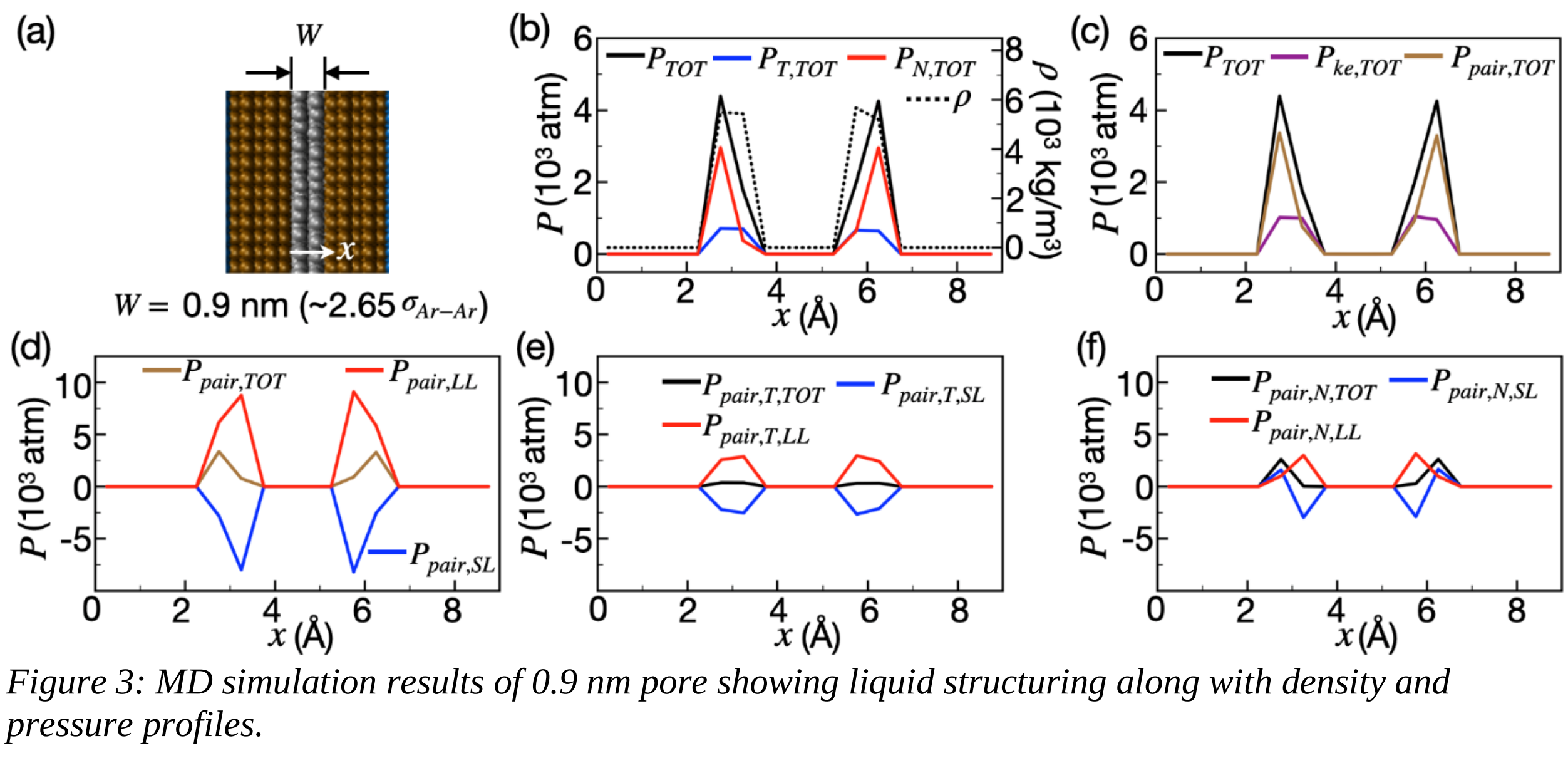

*Figure 3: MD simulation results of 0.9 nm pore showing liquid structuring along with density and pressure profiles.*

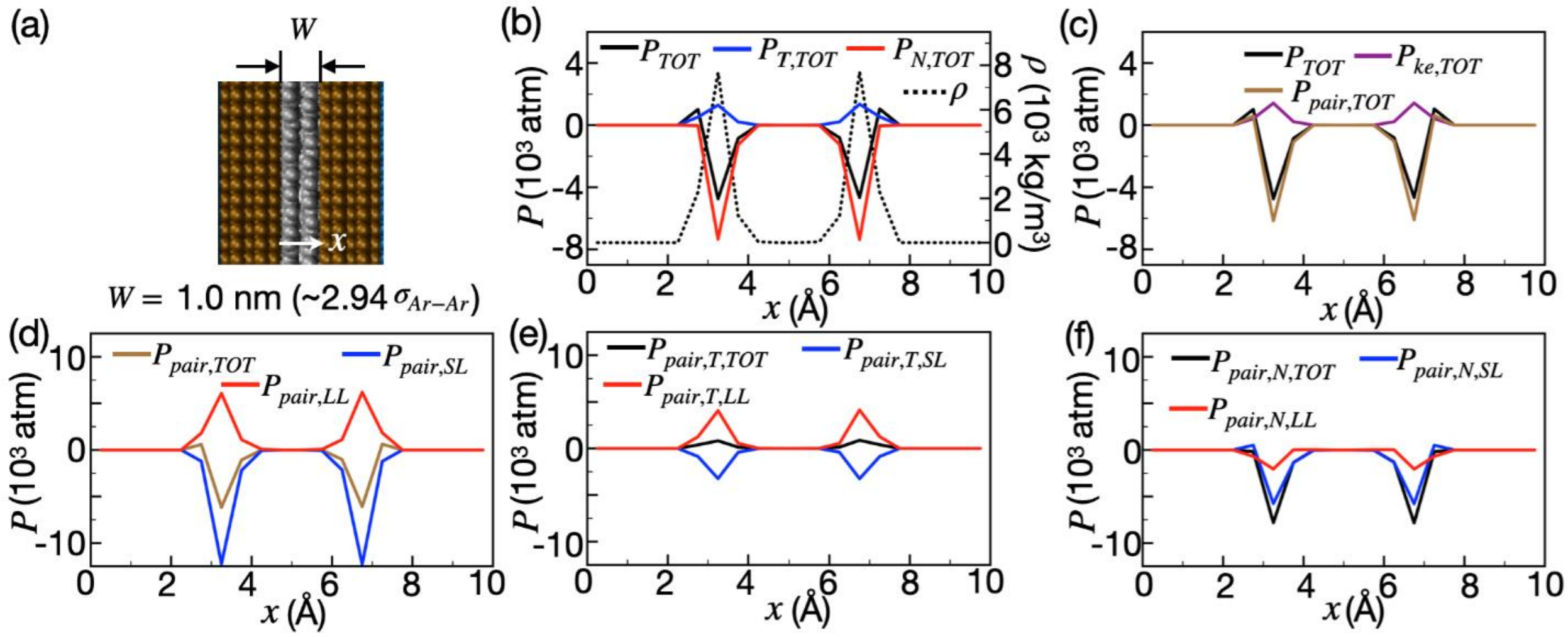

*Figure 4: MD simulation results of 1 nm pore showing liquid structuring along with density and pressure profiles.*

As intermolecular interactions account for the pairwise pressure $P_{pair}$, the force-separated pressures were compared in terms of $P_{pair}$ (Figs. 3d-f and 4d-f). Solid-liquid pairwise pressure $P_{pair,SL}$ is negative for both cases and is in the same order of magnitude as liquid-liquid pairwise pressure $P_{pair,LL}$ which is positive for both cases (Figs. 3d and 4d). The $P_{pair,TOT}$ pressure is either negative or positive depending on which contribution dominates ($P_{pair,SL}$ or $P_{pair,LL}$). Further, $P_{pair,SL}$ can possibly be turned positive by minimally decreasing the pore size (i.e. having the same number of atoms in a reduced volume). However, the liquid atoms might be pushed out of the pore before that occurs, resulting in one liquid layer with negative pressure (described earlier as the periodic pressure oscillation). Dividing the pressure components further, liquid-liquid tangential pairwise pressure $P_{pair,T,LL}$ (Figs. 3e and 4e) is positive in both cases due to the layering effect while normal pairwise pressure $P_{pair,N,LL}$ is affected by the distance between two liquid layers and is dependent on pore size. $P_{pair,N,LL}$ (Figs. 3f and 4f) is positive in 0.9 nm pore and negative in 1.0 nm pore. It can be made more negative by minimally increasing the pore size, but a third liquid layer might form before the negative $P_{N,LL}$ overwhelms $2 \times P_{T,LL}$ in turn causing a jump in pore pressure to reach the peak of next pressure oscillation period. Thus, the occurrence of the negative pressure in nanopore is mostly governed by $P_{pair,SL}$.

Larger size nanopores of 2 nm and 3 nm were also simulated where more than two layers were present; identical observations were found for the layer adjacent to the surface (named as 1st layer) obtained in the 2 nm pore (~$5.88\sigma_{Ar-Ar}$, 5 layers in total, pore pressure -172.40 ± 18.92 atm) and the 3 nm pore (~$8.82\sigma_{Ar-Ar}$, 6 layers with bulk, pore pressure 62.92 ± 15.24 atm); please see supplementary material for data profiles in 2 nm and 3 nm pores. One distinct observation for the layers beyond the 1st layer is that the liquid-liquid pairwise pressure $P_{pair,LL}$ dominates as solid-liquid intermolecular forces exponentially diminish away from the surface. Next, we focus only on the 1st layer.

## B. Structure of 1st Layer in Nanopores

Liquid layering occurs at the solid-liquid interface as the solid-liquid interaction overwhelms liquid-liquid interaction; thus the surface prevents the liquid atoms from moving freely and forces them to align as per the configuration of the surface atoms. Here we define these layers by the sequence of their occurrence from the surface, which can be identified from density profile. Figure 5 shows a typical density profile in 3.0 nm pore. Only half of the domain was analyzed due to the symmetrical nature of the pore. The number of layers depends on the pore size, for e.g., only the 1st layer exists in 0.9 nm and 1.0 nm pores. The 1st layer is defined as the range from the first non-zero point in density profile to the first non-zero minimum point; the range from the first to the second non-zero minimum point is defined as the 2nd layer (Fig. 5). The structure of these layers is due to the competing effects of solid-liquid and liquid-liquid interactions, and is a key factor in determining the pressure in the nanopore. In order to analyze the structure of the 1st layer, we introduce a distribution function $g(x) = \frac{N(x)}{N_{layer}}$, where $N_{layer}$ is the number of atoms in the region between $x - \frac{1}{2}dx$ and $x + \frac{1}{2}dx$, $dx$ is the spatial resolution and chosen as 0.001 nm, and $N_{layer}$ is the number of atoms in the entire layer. As shown in Fig. 6, the 1st layer is much more condensed in 0.9 nm pore than that in other cases. The 1st layer ranges between 0.25 nm and 0.38 nm away from the surface in 0.9 nm pore; while it ranges from 0.25 nm to 0.45 nm in all other three cases. Such slight differences can result in significant pressure variation from positive to negative values.

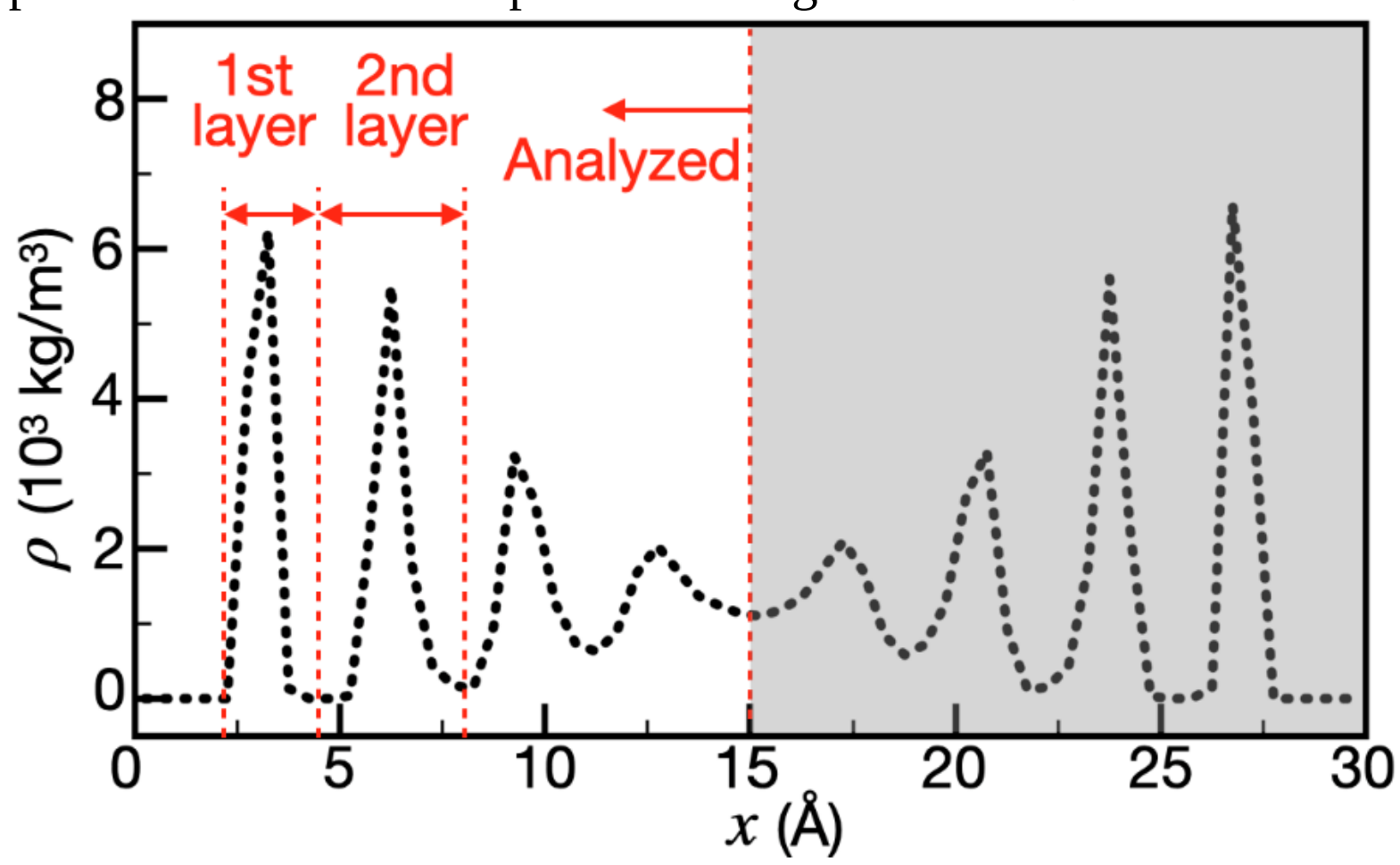

*Figure 5: Definition of the structured layers in nanopores.*

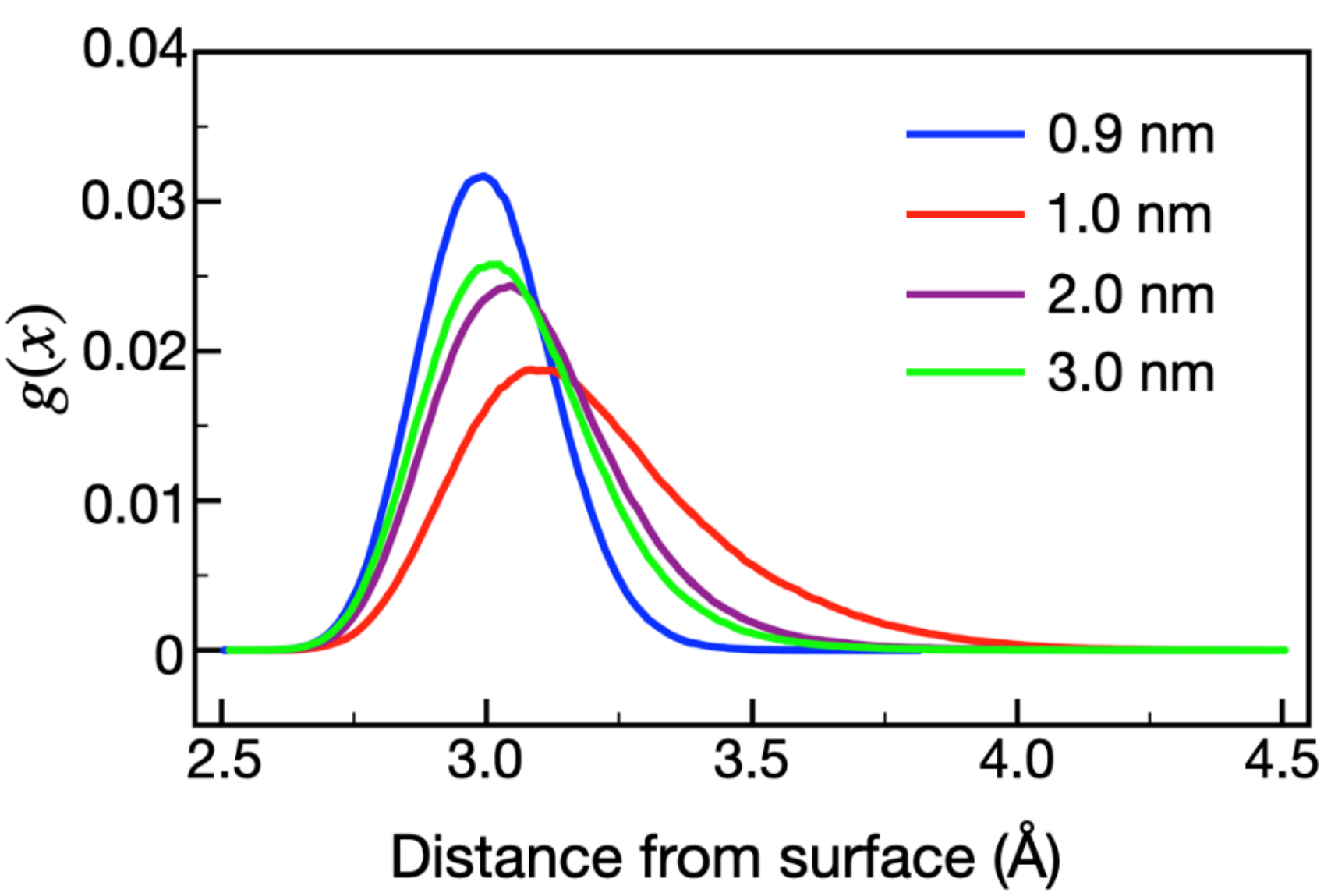

*Figure 6: Distribution function of the first layer in nanopores for varying pore sizes.*

As the surfaces forming the nanopores are infinite into the plane, the solid-liquid force experienced by the liquid atoms in y-z plane (parallel to the surface) is ignorable compared to that in x-y plane (normal to the surface). Thus, we divided the 1st layer into positive and negative force regions based on the force field in x direction (Fig. 7). The positive force region was further divided into strong positive and weak positive regions; while the negative force region was divided into three sub-regions: a strong negative region, where the attractive force is maximum, sandwiched between two weak negative regions. We chose 80% of the maximum absolute value of the negative force as the critical value, above which strong positive or strong negative regions were defined. It should be noted that the force field differs with varying pore sizes. Figure 7a represents the force field for 3 nm pore and Fig. 7b shows the proportion of atoms in each region for all four pore sizes (0.9 nm, 1 nm, 2 nm and 3 nm). Due to the narrow atom distribution in 0.9 nm pore, with a peak closer to the surface compared to other cases (Fig. 6), ~75% of the atoms are located in the positive region, causing a positive average force on a single liquid atom in the 1st layer ($4.04 \times 10^{-14}$ N), and thus a positive local pressure. The proportions of atom in negative force region are ~75%, ~59%, and ~53%, for 1.0 nm, 2.0 nm, and 3.0 nm pores respectively, leading to a negative force on a single atom in the 1st layer ($-3.21 \times 10^{-14}$ N, $-1.82 \times 10^{-14}$ N, and $-0.74 \times 10^{-14}$ N respectively), and thus a negative local pressure.

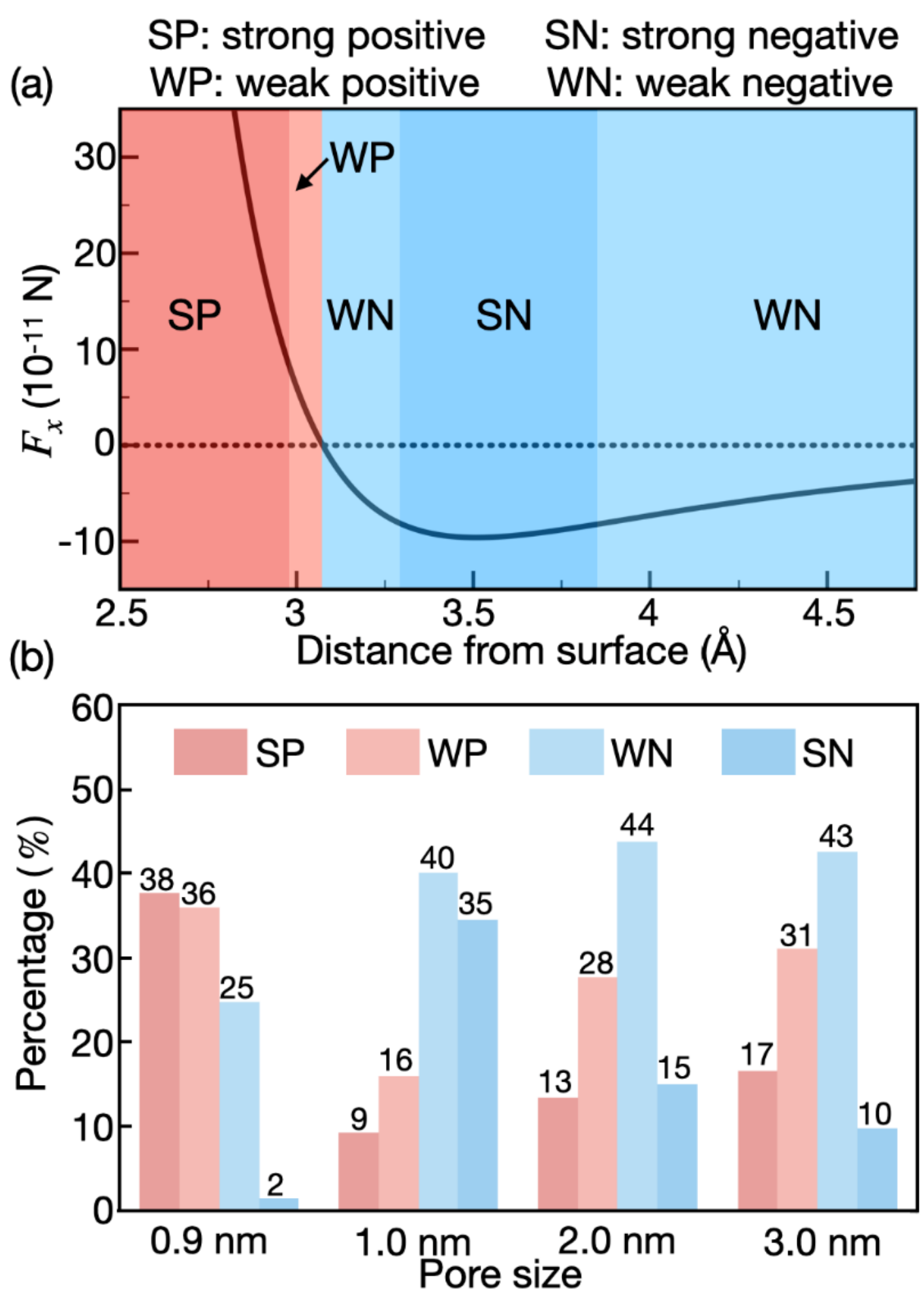

*Figure 7: (a) Regions defined in the 1st layer based on force field; and (b) the proportion of atoms in each region for different pore sizes.*

### C. Effect of Bulk Pressure

Next, we discuss the effect of bulk liquid pressure on pore liquid pressure. In MD simulations, the bulk pressure was tuned by adjusting the number of atoms in the system by removing or adding atoms in the liquid/vapor interface and vapor phase only (please see supplementary material for detail). After that, the simulation was run for at least 3 ns (600,000 steps) for equilibrium. Typically, it takes less than 1 ns to reach the equilibrium state.[39] Interestingly, the pressure in 3 nm pore increased with increasing bulk pressure, with a constant pressure difference $dP = P_{pore} - P_{bulk}$ (Fig. 8a). Constant $dP$ was also observed for 4 nm and 5 nm pores. However, for 2 nm, the pore pressure was constant even though the bulk pressure kept increasing, thus resulting in decreasing $dP$ (Fig. 8b); the same observations were also found for 1 nm pore. Thus, although confinement effect shows up for all nanopores in current work, pressure in the small pores (< 2 nm) is found to be independent of bulk.

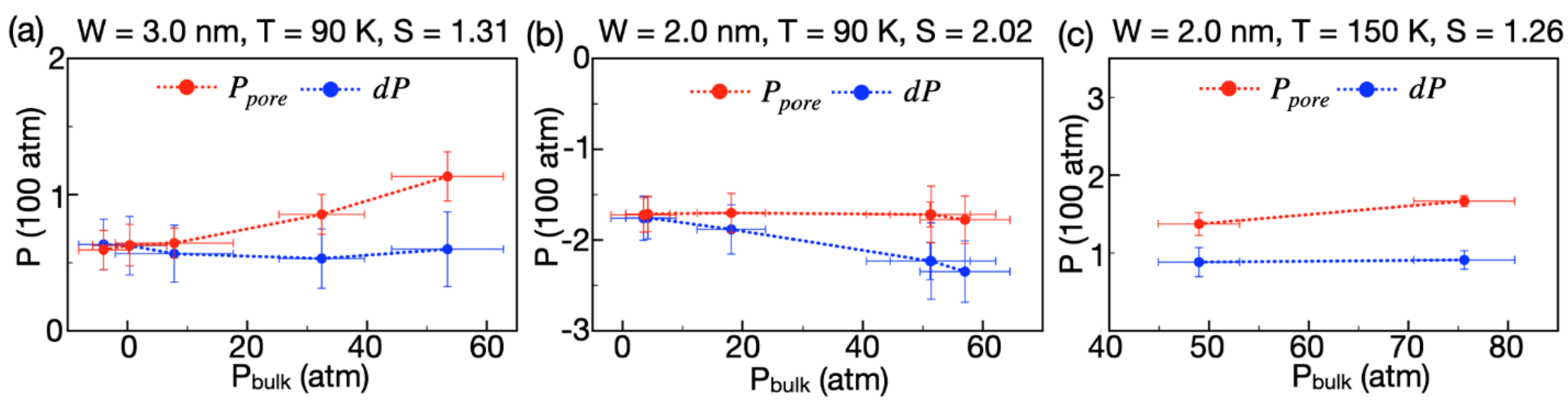

*Figure 8: Effect of bulk pressure on liquid pore pressure. (a) 2.0 nm pore at 90 K with surface influence number S = 2.02, (b) 3.0 nm pore at 90 K with S = 1.31, and (c) 2.0 nm pore at 150 K with S = 1.26.*

In the nanopores, the essence of confinement effect is the competition between solid-liquid and liquid-liquid interactions. The solid surface intends to align liquid atoms to a fixed configuration while the liquid atoms prefer relative freedom and random configuration similar to bulk. The ability of a liquid atom to escape the control of the surface is determined by its thermal energy. With this basis, we introduce a surface influence number, $S$, as a measure of the degree of liquid confinement. It is defined as the ratio between the total surface potential energy experienced by liquid atoms and their kinetic energy (Eq. 4).

$$S = \frac{|\overline{\Phi}_s|}{E_k} = \frac{\int_0^W N(x)|\phi(x)|dx}{\frac{3}{2}k_B T \cdot \int_0^W N(x)dx} \quad (4)$$

where $\overline{\Phi}_s$ is total surface potential energy felt by liquid atoms; $E_k$ is the kinetic energy; $W$ is the nanopore width; $N(x)$ and $\phi(x)$ are the number of atoms and the surface potential energy at location $x$ respectively; $k_B$ is Boltzmann constant; $T$ is temperature.

Due to the discontinuous nature in MD simulations, $S$ is estimated by summing the properties ($N(x)$ and $\phi(x)$) in 1D parallel-to-surface 0.05 nm thick bins in nanopore (Eq. 5). Only the atoms in the center region were included (box #1 in Fig. 1). From the definition, it is expected the solid-liquid interaction dominates when $S \gg 1$, indicating a bulk-independent pore pressure; while a bulk-dependent pore pressure will occur when $S \approx 1$ as the thermal energy will be similar to surface potential energy.

$$S \approx \frac{\sum_{i=1}^{N_b} N(i)|\phi(i)|}{\frac{3}{2}k_B T \cdot \sum_{i=1}^{N_b} N(i)} \quad (5)$$

where $N_b$ is the total number of the bins in nanopore; $N(i)$ is the number the atoms in $i$th bin; $\phi(i)$ is the average surface energy of a single atom in $i$th bin.

Figure 9 plots $S$ values obtained from MD simulation at 90 K versus nanopore width. $S$ number decreases with increased nanopore width based on a power law obtained from curve fitting $S = \frac{4.52}{W^{1.13}}$, with $R^2$ value of 0.9991. For the 3 nm pore where pore pressure increased with increasing bulk pressure, $S$ was estimated to be 1.31, thus implying ~30% higher potential energy relative to kinetic energy is not sufficient to independently dominate liquid properties in the pore. It should be noted that $S < 1$ does not necessarily imply weak confinement effects as the solid-liquid interaction is still significant; for example, in the 5 nm pore where $S$ = 0.78, the pore pressure is still as high as 56.60 ± 12.39 atm with a bulk pressure of 2.72 ± 3.24 atm. To further verify the dependence of pore pressure to bulk pressure as characterized by $S$ number, an additional set of MD simulations were run for liquid confined in 2 nm pore at 150 K. Due to the increased fluid thermal energy, $S$ number decreased to 1.26, similar to that for 3.0 nm pore at 90 K. As expected, the pore pressure increased with increasing bulk pressure, resulting in a constant dP (Fig. 8c).

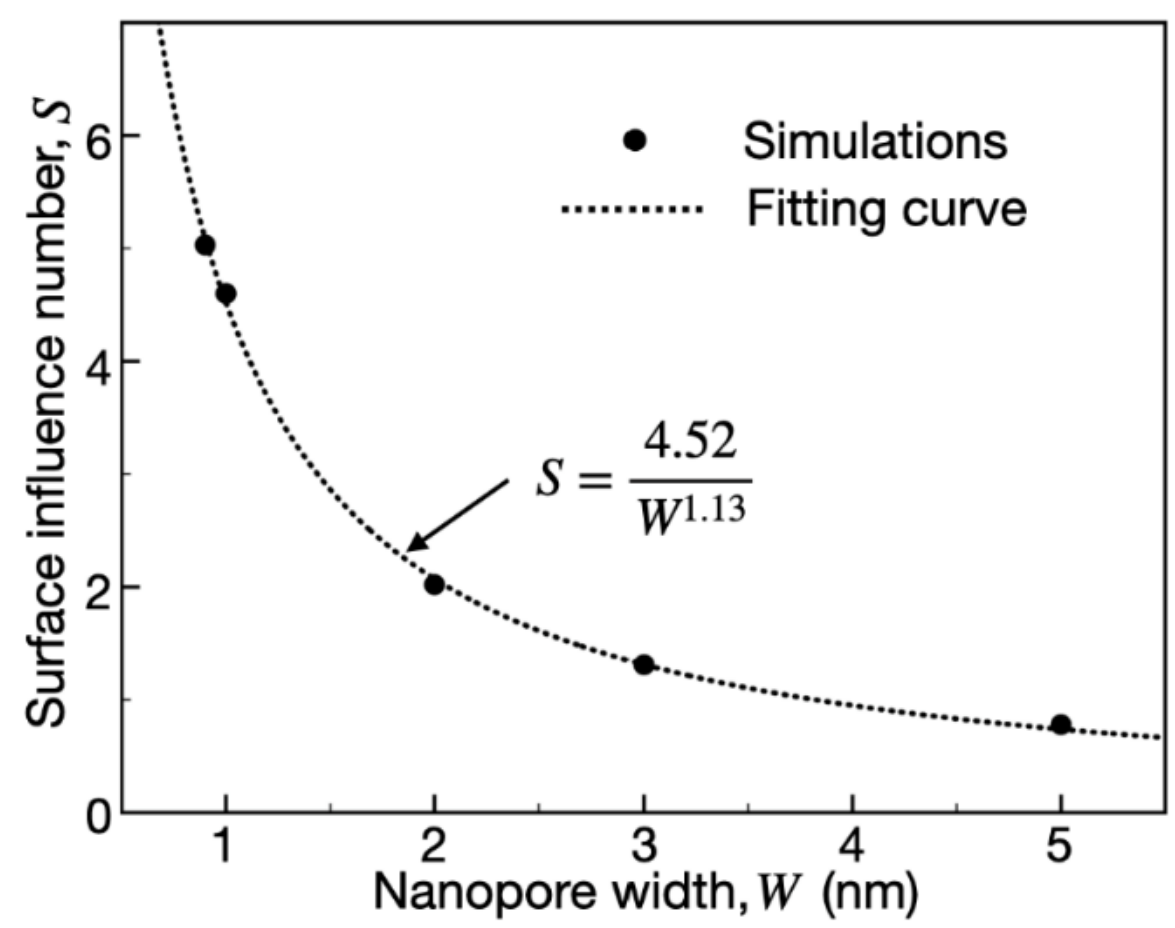

*Figure 9: Surface influence number variation with nanopore width at 90 K.*

## IV. CONCLUSIONS

We report a molecular dynamics simulations study on the confinement effects of argon liquid in hydrophilic nanopores while being connected to bulk liquid. The pressure contributions from solid and liquid atom groups were separated and quantified based on a force separation scheme. The equilibrium pressure in the pore was found to be sensitive to the pore size, and can be tuned to obtain absolute positive or absolute negative values. The contribution from solid-liquid interaction dominated the pressure in the first liquid layer adjacent to the surface while the liquid-liquid interaction dominated the pressure beyond that layer.

The structure of the first liquid layer plays an important role on the pore pressure, especially in pores where only two such layers are present. In a narrow pore, the first layer atoms were located in a small range of distance from the surface causing most liquid atoms to be in the positive force region, thus resulting in positive pressure for the entire layer. On the other hand, for a pore just a little wider (by 0.1 nm), the first liquid layer atoms were further away from the surface causing most atoms to be located in the negative force region, thus resulting in negative liquid pressure.

The effect of bulk pressure on pore pressure was found to be dependent on pore size as well as system temperature. The confinement effect is essentially a competition between solid-liquid and liquid-liquid interactions. A surface influence number $S$ was introduced to quantitatively characterize the degree of the confinement. $S$ number decreased with increasing pore size following a power law function for a constant temperature system. For systems with small $S$ number (less than ~1.3), the pore pressure was affected by the bulk pressure, while for systems with large $S$ number (larger than ~2) the pore pressure was independent of bulk pressure.

## SUPPLEMENTARY MATERIAL

See supplementary material for details on the fluid state in MD simulations, tuning of bulk pressure, verification of force separation scheme, and pressure tensor data for 2 nm and 3 nm pores.

## ACKNOWLEDGEMENT

This material is based upon work supported by, or in part by, the Office of Naval Research under contract/grant no. N000141812357.

## DATA AVAILABILITY STATEMENT

The data that support the findings of this study are available from the corresponding author upon reasonable request.